\documentclass[12pt]{article}
\usepackage{eurosym}
\usepackage{graphicx}
\usepackage{mathptmx,mathrsfs}
\usepackage{amsfonts,amsmath,amssymb,amsthm}
\usepackage{indentfirst}
\usepackage{cite}
\usepackage{hyperref}
\usepackage{enumitem}

\DeclareMathAlphabet{\mathcal}{OMS}{cmsy}{b}{n}

\numberwithin{equation}{section}

\theoremstyle{definition}

\begin{document}

\title{Causal approach for the electron-positron scattering in the
generalized quantum electrodynamics}
\author{R. Bufalo$^{1,2}$\thanks{%
rbufalo@ift.unesp.br}~, B.M. Pimentel$^{2}$\thanks{%
pimentel@ift.unesp.br}~, D.E. Soto$^{2}$\thanks{%
danielsb@ift.unesp.br}~ \\
\textit{{$^{1}${\small Department of Physics, University of Helsinki, P.O. Box 64}}}\\
\textit{\small FI-00014 Helsinki, Finland}\\
\textit{{$^{2}${\small Instituto de F\'{\i}sica Te\'orica (IFT), UNESP,  Universidade Estadual Paulista (UNESP)}}} \\
\textit{\small Rua Dr. Bento Teobaldo Ferraz 271, Bloco II Barra Funda, CEP
01140-070 S\~ao Paulo, SP, Brazil}\\
}
\maketitle
\date{}

\begin{abstract}
In this paper we study the generalized electrodynamics contribution to the electron-positron scattering process
$e^{-}e^{+}\rightarrow e^{-}e^{+}$, i.e., Bhabha scattering. Within the framework of the standard model and for energies
larger than the electron mass, we calculate the cross section for the scattering process. This quantity is usually calculated in the framework of Maxwell electrodynamics and (for phenomenological reasons) is corrected by a cutoff parameter. On the other hand, by considering generalized electrodynamics instead of Maxwell's,
we show that the Podolsky mass plays the part of a natural cut-off parameter for this scattering process. Furthermore, by using experimental data on Bhabha scattering we estimate its lower bound. Nevertheless, in order to have a mathematically well-defined description of our study we shall present our discussion in
the framework of the Epstein-Glaser causal theory.
\end{abstract}

\newpage

\tableofcontents

\section{Introduction}

Perturbative Quantum Electrodynamics (QED) is a gauge theory that represents a remarkable computational success.
For example, one may cite its impressive accuracy regarding the measurement of the anomalous magnetic moments of the
electron and the muon \cite{Kinoshi}. However, it is a well-known fact that the standard model of particle physics is
nothing more than an effective theory \cite{WeingEff}, although its energy range is still a matter of  discussion \cite{ref55}. Because of this, there is room for different theoretical proposals for describing, for instance, the electromagnetic field; thus, if we subscribe to the standard lore of effective field theories, all possible terms allowed by the symmetries of the theory ought to be included. An interesting group of such effective theories are the higher-order derivative (HD) Lagrangians \cite{Ostro}. They were initially proposed as an attempt to achieve a better ultraviolet behavior and renormalizability properties of physically relevant models. Moreover, in electromagnetism it is known that the Maxwell Lagrangian depends, at most, on first-order derivatives. However, one may add a second-order term in such a way that all original symmetries are preserved. In fact, it was proven in Ref.\cite{ref29} that such a term is unique when one requires the preservation of the theory's linearity and Abelian $U(1)$ and Lorentz symmetries. As a result, we have the generalized electrodynamics Lagrangian introduced by Bopp \cite{Bopp} and Podolsky and Schwed \cite{BorisP}. Moreover, an important feature of generalized quantum electrodynamics (GQED) is that -- in the same way that the Lorenz condition is a natural gauge condition for the Maxwell electrodynamics -- it has a counter-part, the so-called generalized Lorenz condition \cite{GalvPim}: $\Omega \left[ A\right] =\left( 1+a^{2}\square \right) \partial ^{\mu }A_{\mu }$. A recent study via functional methods has shown that the electron self-energy and vertex part of GQED are both ultraviolet finite at $\alpha $ order \cite{RBufBMP1}, as well as the theory's renormalizability \cite{RBufBMP}.

Moreover, despite the radiative functions previously evaluated, there are still some interesting scenarios where
one may search for deviations from standard physics that are rather important, such as the scattering of standard model particles \cite{Mandels}. In particular, we may cite the study of Moeller scattering \cite{Moeller}, $e^{-}e^{-}\rightarrow e^{-}e^{-}$, and Bhabha scattering \cite{Bhabha}, $ e^{-}e^{+}\rightarrow e^{-}e^{+}$, as offering some particularly interesting possibilities due to their large cross section, which lead to very good statistics. On the other hand, the modern linear electron collider allows for experiments with high-precision measurements. Another important experiment is the annihilation process $e^{+}e^{-}$ producing a pair of leptons; in particular, we have the muon pair production $e^{-}e^{+}\rightarrow \mu ^{-}\mu ^{+}$ and tau pair production $e^{-}e^{+}\rightarrow \tau ^{-}\tau ^{+}$. We can cite studies in the literature that analyzed deviations due to Lorentz-violating effects of the cross section in the electron-positron annihilation \cite{ref20}. There has also been a recurrent discussion on improving two-loop calculations for Bhabha scattering using contributions from QED \cite{2loopqed}.

Bhabha scattering is one of the most fundamental reactions in QED processes, as well as in phenomenological
studies in particle physics. Also, it is particularly important mainly because it is the process employed
in determining the luminosity at $e^{+}e^{-}$ colliders. At colliders operating at c.m. energies of
$\mathcal{O}\left( 100~\text{GeV}\right) $ the relevant kinematic region is the one in which the angle between
the outgoing particles and the beam line is only about a few degrees. In these regions the Bhabha scattering
cross section is comparatively large and the QED contribution dominates. Since the luminosity value is measured
with very high accuracy \cite{Liminosity}, it is necessary to have a precise theoretical calculation of the value for
the Bhabha scattering cross section in order to keep the error in the luminosity small.

Despite the incredible match between theoretical and experimental values in QED, there are some particularly
intriguing discrepancies between the QED results and measurements (even when electroweak and strong interaction effects are included). These discrepancies are on the order of one standard deviation, such as those in the 1S ground-state Lamb shift of hydrogenic atoms \cite{ref30} and in the magnetic moment of the muon \cite{ref31}. These facts give us, in principle, a window of possibilities for proposing a modification to the QED vertex and/or the photon propagator; thus, we could in principle calculate a lower limit for the mass of a massive "photon" (GQED gives both massless and massive propagating modes). Nevertheless, since generalized quantum electrodynamics is a good alternative for describing the interaction between fermions and photons, we shall consider this theory in order to calculate the first-order correction to the usual QED differential cross section for Bhabha scattering. For this purpose we shall consider the framework of the perturbative causal theory of Epstein and Glaser \cite{EpsGlas}, specifically its momentum-space form developed by Scharf et al \cite{Scharf}.

Therefore, in this paper, we calculate within the framework of the Epstein-Glaser causal theory
the contribution of  generalized electrodynamics to the Bhabha scattering cross section.
This paper is organized as follows. In Sec.\ref{sec:1}, by means of the theory of distributions,
we introduce the analytic representation for the positive, negative and causal propagators; moreover, we introduce an alternative
gauge condition that is different from the generalized Lorentz condition. In Sec.\ref{sec:2} we obtain a well-defined (Feynman) electromagnetic propagator in the causal approach. \footnote{This is simply called the electromagnetic propagator; however, since we introduce several propagators in this paper, we shall adopt the notation of Ref.\cite{JDBjor}.} Finally, in Sec.\ref{sec:3} we calculate the GQED correction for Bhabha scattering, and by
using the experimental data for this process we determine a lower bound for the Podolsky mass.
In Sec. \ref{sec:5} we summarize the results, and present our final remarks and prospects.

\section{Analytic Representation for Propagators}

\label{sec:1}

In order to develop the analytic representation for propagators we shall consider the Wightman formalism \cite{Wightman}.
This axiomatic approach guarantees that general physical principles are always obeyed. To formulate the analytic
representation, we start by discussing the free scalar quantum field in this formalism. \footnote{Actually, we will
briefly review its development, since a detailed discussion can be found in Ref.\cite{ref28}}

The free scalar field, $\phi $, is a general distributional solution of the
Klein-Gordon-Fock equation $\left( \square +m^{2}\right) \phi =0$. The whole theory is formulated in terms of this field, and is understood as an operator-valued distribution defined on the Schwartz space, $\mathcal{J}\left(\mathbb{R}^{4}\right) $. In this space it is possible define the Fourier transformation of the scalar field, $\hat{\phi}\left( k\right) $. Thus, if $ f\in \mathcal{J}\left(\mathbb{R}^{4}\right)$ the distribution $ \phi $ is defined as $\phi \left[ f\right] \overset{or}{=}\left\langle \phi ,f\right\rangle $, and $\hat{\phi}\left( k\right) $ is defined as
\begin{equation}
\left\langle \hat{\phi},\check{f}\right\rangle =\left\langle \phi
,f\right\rangle =\int dk\hat{\phi}\left( k\right) \check{f}\left( k\right) ,
\end{equation}%
where $\check{f}$ is the inverse Fourier transformation of $f$, which also belongs to $\mathcal{J}\left(\mathbb{R}^{4}\right) $.

In the Wightman formalism the field $ \phi $ generates the full Hilbert space from the invariant vacuum $\left\vert \Omega \right\rangle $. Moreover, we can formally split the field into the positive- (PF) and negative-frequency (NF)
components, defined in the distributional form as
\begin{align}
\phi ^{\left( \pm \right) }\left[ f\right] \left\vert \Omega \right\rangle
=\int d^{4}k\theta \left( k_{0}\right) \delta \left( k^{2}-m^{2}\right)
\tilde{a}\left( k\right) \hat{f}\left( \pm  k\right) \left\vert \Omega\right\rangle ,  \label{1.1}
\end{align}%
where $\phi ^{\left( +\right) }$ is the positive and $\phi ^{\left(-\right) }$ is the negative part of the
field. By the \textit{spectral condition}, we obtain that we do not have components in
$k\in V^{-}\rightarrow -k\in V^{+}$; we then see that the part associated to $\hat{f}\left( -k\right) $
must be zero, while the part associated to $\hat{f}\left( k\right) $ must be nonzero. These conditions are satisfied only if%
\begin{align}
\tilde{a}\left( k\right) \left\vert \Omega \right\rangle =0, \qquad
\tilde{a}^{\dag }\left( k\right) \left\vert \Omega \right\rangle \neq 0.
\end{align}%
From these conditions it follows that the operators $\tilde{a}\left( k\right) $ and $\tilde{a}^{\dag }\left( k\right) $
are interpreted as the operators of annihilation and creation, respectively.

In this formalism the central objects are the so-called Wightman functions. They are defined as the vacuum
expectation values (VEVs) of a product of fields. For instance, the two-point Wightman function for scalar fields is given by%
\begin{align}
W_{2}\left( x_{1},x_{2}\right) \equiv  \left\langle \Omega \right\vert \phi
\left( x_{1}\right) \phi \left( x_{2}\right) \left\vert \Omega \right\rangle =\left( 2\pi \right) ^{-2}\int d^{4}k\hat{W}%
_{2}\left( k\right) e^{-ik\left( x_{1}-x_{2}\right) },
\end{align}%
where $\hat{W}_{2}\left( k\right) $ is the two-point Wightman function in momentum space. Of course they
are not functions in the strict sense, but rather distributions defined on $\mathcal{J}\left(\mathbb{R}^{4}\right) $. Moreover, we have that the Wightman function obeys the same equation as that for the free field. Hence, as a consequence of the spectral condition, one can find that the two-point Wightman function in momentum space is given by
\begin{equation}
\hat{W}_{2}\left( k\right) =\frac{1}{2\pi }\theta \left( k_{0}\right) \delta
\left( k^{2}-m^{2}\right) .  \label{1.4}
\end{equation}
With the necessary physical concepts and tools in hand, we shall now introduce the analytic representation
for the propagators. In order to elucidate the content we shall discuss the case of scalar fields first.

\subsection{Analytic representation of the PF and NF propagators}

Since the fundamental propagators are linear combinations of the PF and NF frequency parts of the propagator, it is rather natural to consider them here in our development. We define the PF propagator by the relation of the contraction between scalar fields:%
\begin{equation}
\overbrace{\phi \left( x\right) \phi \left( y\right) }\equiv \left[ \phi
^{\left( -\right) }\left( x\right) ,\phi ^{\left( +\right) }\left( y\right) %
\right] =-iD_{m}^{\left( +\right) }\left( x-y\right) .
\end{equation}%
Moreover, for a normalized vacuum, we have that the PF and NF propagators can be written as follows%
\begin{equation}
D_{m}^{\left( \pm \right) }\left( x-y\right) =i\left\langle \Omega \left\vert
\left[ \phi ^{\left( \mp \right) }\left( x\right) ,\phi ^{\left( \pm \right)
}\left( y\right) \right] \right\vert \Omega \right\rangle .
\end{equation}%
Now, by using the properties of the positive and negative parts of the field, Eq.\eqref{1.1},
we may find the relation between the PF and NP propagators, as well as the relation to the Wightman function%
\begin{equation}
D_{m}^{\left( -\right) }\left( x-y\right) =-D_{m}^{\left( +\right) }\left(
y-x\right) =-iW_{2}\left( y-x\right) .
\end{equation}%
Hence, with the above results we can make use of Eq.\eqref{1.4} to obtain the PF and NF propagators written in momentum space,
\begin{equation}
\hat{D}_{m}^{\left( \pm \right) }\left( k\right) =\pm \frac{i}{2\pi }\theta
\left( \pm k_{0}\right) \delta \left( k^{2}-m^{2}\right) =%
\frac{i}{2\pi }\frac{\delta \left( k_{0}\mp \omega _{m}\right) }{k_{0}\pm
\omega _{m}},  \label{1.5}
\end{equation}%
where $\omega _{m}=\sqrt{\vec{k}^{2}+m^{2}}$ is the frequency. To write down the analytic representation of the scalar propagator, we should remark that the second equality in Eq.\eqref{1.5} must be understood as distributions in $k_{0}$. Thus, after using the definition of the Dirac $\delta$-translated distribution \eqref{B.1} and Cauchy's integral theorem, we find that the propagators $\hat{D} _{m}^{\left( \pm \right) }$ can be defined by the following analytic representation \cite{ref28}:
\begin{equation}
\left\langle \hat{D}_{m}^{\left( \pm \right) },\varphi \right\rangle =\left(2\pi \right) ^{-2}\oint
\limits_{c_{\pm }}\frac{\varphi \left( k_{0}\right) }{k_{0}^{2}-\omega _{m}^{2}}dk_{0},  \label{1.6}
\end{equation}%
where $c_{+\left( -\right) }$ is a counterclockwise closed path that contains only the positive (negative)
poles of the Green's function $\hat{g}\left( k\right) =\frac{1}{k_{0}^{2}-\omega _{m}^{2}}$.

Nevertheless, we should emphasize that Eq.\eqref{1.6} may be generalized for any free field. Thus, if $ \hat{D}^{\left( \pm \right) } $ are the PF and NF propagators associated to an arbitrary field $ A $, then, its analytic representations are given by
\begin{equation}
\left\langle \hat{D}^{\left( \pm \right) },\varphi \right\rangle =\left(
2\pi \right) ^{-2}\oint\limits_{c_{\pm }}\hat{G}\left( k\right) \varphi\left( k_{0}\right) dk_{0}.  \label{1.7}
\end{equation}%
where $c_{+\left( -\right) }$ is a counterclockwise closed path that contains only the positive (negative) poles on the $ k_{0}$-complex plane of the Green's function $\hat{G}\left( k\right) $ associated to the free-field equation of the field $ A $.

Moreover, since the PF and NF propagators are distributional solutions of the free-field equations, any linear
combination of these is also a solution; for example, we may define the \emph{causal} propagator distributional solution
\begin{equation}
\hat{D}\left( k\right) =\hat{D}^{\left( +\right) }\left( k\right) +\hat{D}%
^{\left( -\right) }\left( k\right) , \label{eq 0.5}
\end{equation}%
and from the spectral condition \cite{Wightman} its support is given as it follows \footnote{The regions $\bar{%
V}^{\pm }$ are the closed forward and backward cones defined as: $\bar{V}^{\pm }=\left\{ x ~| ~ x^{2}\geq 0,\quad \pm x_{0}\geq 0\right\} .$}:
\begin{equation}
\text{Supp}~\hat{D}\left( k\right) =\text{Supp}~\hat{D}^{\left( +\right) }\left( k\right)
\cup \text{Supp}~\hat{D}^{\left( -\right) }\left( k\right) =\bar{V}^{+}\left(k\right) \cup \bar{V}^{-}\left( k\right) .
\end{equation}%

\subsection{The PF and NF electromagnetic propagators}

\label{sec:2.2}

The dynamics of the generalized electromagnetic theory is governed by the Lagrangian density as follows \cite{Bopp,BorisP}:
\begin{equation}
\mathcal{L}_{P}=-\frac{1}{4}F_{\mu \nu }F^{\mu \nu }+\frac{a^{2}}{2}\partial
_{\mu }F^{\mu \sigma }\partial ^{\nu }F_{\nu \sigma },
\end{equation}%
where $F_{\mu \nu }=\partial _{\mu }A_{\nu }-\partial _{\nu }A_{\mu }$ is the usual electromagnetic tensor field
and $a$ is the free Podolsky parameter with the dimension of length. This Lagrangian is invariant under $U(1)$ gauge and Lorentz transformations. Usually the gauge-fixing procedure is performed by adding a Lagrange multiplier to the
Lagrangian density if we consider the Lorenz condition $\left( \partial ^{\mu }A_{\mu }\right) ^{2}$. However,
if we instead consider the generalized Lorenz condition \cite{GalvPim,RBufBMP}, we shall add the term $\left[ \left(
1+a^{2}\square \right) \partial ^{\mu }A_{\mu }\right] ^{2}$. This condition was proven to be the natural choice
for generalized electrodynamics; however, it increases the order of the field equation. Nevertheless, in order
to preserve the order of the field equation, we may consider a third choice, namely, we add an alternative
gauge-fixing term to the Lagrangian $\left( \partial .A\right) \left( 1+a^{2}\square \right) \left(\partial .A\right) $ \cite{ref52}, which is called the \emph{nonmixing gauge} and it is related to a pseudodifferential operator \cite{Claus}. So, in this gauge condition, the total Lagrangian density is given by
\begin{align}
\mathcal{L}_{P}=-\frac{1}{4}F_{\mu \nu }F^{\mu \nu }+\frac{a^{2}}{2}%
\partial _{\mu }F^{\mu \sigma }\partial ^{\nu }F_{\nu \sigma }-\frac{1}{%
2\xi }\left( \partial .A\right) \left( 1+a^{2}\square \right) \left(\partial .A\right) ,
\end{align}%
where $\xi $ is the gauge-fixing parameter. From this equation we can find the equation of motion,%
\begin{equation}
\mathcal{E}_{\mu \nu }\left( \partial \right) A^{\nu }\equiv \left(1+a^{2}\square \right) \left[ \left( \square g_{\mu \nu }-\partial _{\mu
}\partial _{\nu }\right) +\dfrac{1}{\xi }\partial _{\mu }\partial _{\nu }\right] A^{\nu }=0.  \label{1.7a}
\end{equation}%
We see that for the choice $\xi =1$, the equation of motion is simply reduced to $\left(
1+a^{2}\square \right) \square A_{\mu }=0$, which clearly indicates the presence of two sectors in the free case: a Maxwell sector,
\begin{equation}
\square A_{\mu }=0,
\end{equation}
and a Proca sector,
\begin{equation}
\left( \square +m_{a}^{2}\right) A_{\mu }=0,
\end{equation}
where $m_{a}=a^{-1}$ is the Podolsky mass. Moreover, in order to determine the analytic representation for the propagator we should first determine the Green's function of Eq.\eqref{1.7a}. In fact, it reads
\begin{align}
\hat{G}_{\mu \nu }\left( k\right) =\left[g_{\mu \nu }-\left( 1-\xi \right) \frac{k_{\mu }k_{\nu }}{m_{a}^{2}}\right]\left( \frac{1}{k^{2}}-\frac{1%
}{k^{2}-m_{a}^{2}}\right) -\left( 1-\xi \right) k_{\mu }k_{\nu }\frac{1}{\left( k^{2}\right) ^{2}}.
\end{align}%
Thus, to find the PF and NF electromagnetic propagators via the analytic
representation \eqref{1.7}, it is convenient to calculate each of the above terms separately:

\begin{enumerate}
\item First we consider the scalar case with Podolsky mass $m_{a}$:%
\begin{equation}
\hat{G}_{m}\left( k\right) =\frac{1}{k_{0}^{2}-\omega _{m_{a}}^{2}}=\frac{1}{%
\left( k_{0}+\omega _{m_{a}}\right) \left( k_{0}-\omega _{m_{a}}\right) },
\end{equation}%
where $\omega _{m_{a}}=\sqrt{\mathbf{k}^{2}+m_{a}^{2}}$. Hence, from Eq.\eqref{1.7} the analytic representation for the PF and NF propagators is written as%
\begin{equation}
\left\langle \hat{D}_{m_{a}}^{\left( \pm \right) },\varphi \right\rangle
=\left( 2\pi \right) ^{-2}\oint\limits_{c_{\pm }}\frac{\varphi \left(
p_{0}\right) }{\left( k_{0}+\omega _{m_{a}}\right) \left( k_{0}-\omega _{m_{a}}\right) }dk_{0}.
\end{equation}%
Using Cauchy's integral theorem, as well as some distributional properties of the Dirac $\delta $ distribution \cite{Gwais}
(see Appendix \ref{sec:B}), we obtain the \emph{PF and NF scalar propagators with mass} $m_{a}$ in momentum space:%
\begin{equation}
\hat{D}_{m_{a}}^{\left( \pm \right) }\left( k\right) =\pm \dfrac{i}{2\pi }%
\theta \left( \pm k_{0}\right) \delta \left( k^{2}-m_{a}^{2}\right) . \label{1.8}
\end{equation}%
In particular, for the massless case they have the form%
\begin{equation}
\hat{D}_{0}^{\left( \pm \right) }\left( k\right) =\pm \dfrac{i}{2\pi }\theta
\left( \pm k_{0}\right) \delta \left( k^{2}\right) .  \label{1.9}
\end{equation}

\item Now we shall consider the Green's function $\hat{G}_{0}^{\prime }\left( k\right) =
\left( k_{0}^{2}-\omega _{0}^{2}\right) ^{-2}$, which is associated to the dipolar massless scalar case \cite{Naka}.
Moreover, from the Eq.\eqref{1.7} the analytic representation for this term is given by%
\begin{equation}
\left\langle \hat{D}_{0}^{\prime \left( \pm \right) },\varphi \right\rangle =\left( 2\pi \right) ^{-2}\oint\limits_{c_{\pm }}\frac{\varphi \left( k_{0}\right) }{\left( k_{0}^{2}-\omega _{0}^{2}\right)^{2}}dk_{0}.
\end{equation}%
Using Cauchy's integral theorem again, and after some algebraic manipulations, we have that:%
\begin{align}
\left\langle \hat{D}_{0}^{\prime \left( \pm \right) },\varphi \right\rangle &=\frac{i}{2\pi }\sum\limits_{j=0}^{1}\left( -1\right) ^{j}\frac{\left(
1+j\right) !}{\left( 1-j\right) !}\frac{\varphi ^{\left( 1-j\right) }\left(
\pm \omega _{0}\right) }{\left( \pm 2\omega _{0}\right) ^{2+j}},
\end{align}%
Hence, from the definition of the translated Dirac $\delta $-distribution Eq.\eqref{B.2} and using the distributional property \eqref{B.3} we obtain the \emph{PF and NF dipolar massless scalar propagators}:
\begin{equation}
\hat{D}_{0}^{\prime \left( \pm \right) }\left( k\right) =\mp \frac{i}{2\pi }%
\theta \left( \pm k_{0}\right) \delta ^{\left( 1\right) }\left(k_{0}^{2}-\omega _{0}^{2}\right) .  \label{1.10}
\end{equation}
\end{enumerate}

Finally, we obtain that the expression for the \emph{PF and NF electromagnetic propagators}
in momentum space is given as
\begin{align}
\hat{D}_{\mu \nu }^{\left( \pm \right) }\left( k\right) =&\left[g_{\mu \nu }-\left( 1-\xi \right) \frac{k_{\mu }k_{\nu }}{
m_{a}^{2}}\right]\left(\hat{D}_{0}^{\left( \pm \right) }\left( k\right) -\hat{D}_{m_{a}}^{\left(\pm \right) }\left( k\right) \right) \notag \\
&-\left( 1-\xi \right) k_{\mu }k_{\nu }%
\hat{D}_{0}^{\prime \left( \pm \right) }\left( k\right) ,  \label{1.14}
\end{align}%
where the propagators $\hat{D}_{m_{a}}^{\left( \pm \right) }$, $\hat{D}_{0}^{\left( \pm \right) }$, and
$\hat{D}_{0}^{\prime \left( \pm \right) }$ are given by Eqs.\eqref{1.8}, \eqref{1.9}, and \eqref{1.10}, respectively.

Before introducing the formal causal method itself, it is rather interesting -- by means of complementarity -- to present
some remarks discussed previously for the scalar case that are suitable for the gauge field as well. With the PF
and NF scalar propagators we may determine the \emph{causal} propagator \eqref{eq 0.5}:%
\begin{equation}
D_{m}\left( x\right) =D_{m}^{\left( +\right) }\left( x\right) +D_{m}^{\left(
-\right) }\left( x\right) . \label{eq 0.5a}
\end{equation}%
The causal propagator can also be split into two important propagators: one that indicates the propagation to the future, and one that indicates propagation to the past. These are the \emph{retarded} and \emph{advanced} propagators, which are related to the causal propagator as follows%
\begin{equation}
D_{m}^{R}\left( x\right) =\theta \left( x_{0}\right) D_{m}\left( x\right) , \quad
D_{m}^{A}\left( x\right) =-\theta \left( -x_{0}\right) D_{m}\left( x\right). \label{1.11}
\end{equation}%
From these very definitions, we obtain another important distributional solution -- the so-called \emph{Feynman}
propagator,
\begin{equation}
D_{m}^{F}\left( x\right) =\theta \left( x_{0}\right) D_{m}^{\left( +\right)
}\left( x\right) -\theta \left( -x_{0}\right) D_{m}^{\left( -\right) }\left(
x\right) ,  \label{1.13}
\end{equation}%
which is related to the vacuum expectation value of time-ordered products.
Moreover, for the scalar case this distribution in momentum-space form can be written as follows:
\begin{equation}
\hat{D}_{m}^{F}\left( k\right) =-\left( 2\pi \right) ^{-2}\lim_{\epsilon
\rightarrow 0^{+}}\frac{1}{k^{2}-m^{2}+i\epsilon },
\end{equation}%
which differs from its Green's function by the imaginary term $i\epsilon $. This addition process is
called the Feynman $i\epsilon $-prescription \cite{JDBjor,BogoIntro} and it is closely related to the
Wick-rotation technique. In general this prescription is given in order to handle the singularities in the propagators.

On the other hand, the causal approach takes into account only the general physical properties of the propagators when they are obtained. For instance, when we insert the scalar Feynman propagator [using the definition from
the retarded or advanced distribution \eqref{1.11}] into Eq.\eqref{1.13}, we obtain that%
\begin{equation}
D_{m}^{F}\left( x\right) =D_{m}^{R}\left( x\right) -D_{m}^{\left( -\right)
}\left( x\right) =D_{m}^{A}\left( x\right) +D_{m}^{\left( +\right) }\left(x\right) .
\end{equation}%
This is not a superfluous equivalence to the Wick rotation. Furthermore, when we separate it into positive and negative parts, we may show that the Feynman propagator has the following causal property: \emph{Only the positive-frequency
solution can be propagating to the future and only the negative-frequency solution can be propagating to the past}.
This general physical property and the general definition of the propagator in
the distributional form are the starting points for the development of the Epstein-Glaser causal
approach, in which no prescription is employed in dealing with the propagator's poles.

In order to determine the electromagnetic propagator in the \emph{nonmixing} gauge, we must first obtain
the expression for the \emph{causal} electromagnetic propagator as a sum of the PF and NF propagators
\eqref{1.14} [and using Eq.\eqref{eq 0.5}],
\begin{align}
\hat{D}_{\mu \nu }\left( k\right) =\left[g_{\mu \nu }-\left( 1-\xi \right)
\frac{k_{\mu }k_{\nu }}{m_{a}^{2}}\right]\left( \hat{D}_{0}\left(
k\right) -\hat{D}_{m_{a}}\left( k\right) \right) -\left( 1-\xi \right)
k_{\mu }k_{\nu }\hat{D}_{0}^{\prime }\left( k\right) .  \label{1.16}
\end{align}%
Moreover, using Eqs.\eqref{1.8}, \eqref{1.9}, and \eqref{1.10}, respectively,
the (causal) terms of Eq.\eqref{1.16} are written as follows:
\begin{gather}
\hat{D}_{0} =\frac{i}{2\pi }sgn\left( k_{0}\right) \delta
\left( k^{2}\right), \quad  \hat{D}_{0}^{\prime } = -\frac{i}{2\pi }sgn\left(k_{0}\right) \delta ^{\left( 1\right) }\left( k^{2}\right),  \quad \hat{D}_{m_{a}} =\frac{i}{%
2\pi }sgn\left( k_{0}\right) \delta \left( k^{2}-m_{a}^{2}\right) . \label{1.16a}
\end{gather}
Finally, we can write this propagator in the configuration space,
\begin{align}
D_{\mu \nu }\left( x\right) =\left[g_{\mu \nu }+\left( 1-\xi \right) \frac{%
\partial _{\mu }\partial _{\nu }}{m_{a}^{2}}\right]\left[ D_{0}\left( x\right)-D_{m_{a}}\left( x\right) \right] +\left( 1-\xi \right) \partial _{\mu
}\partial _{\nu }D_{0}^{\prime }\left( x\right)  ,  \label{1.15}
\end{align}%
where $D_{0}\left( x\right) $, $D_{m_{a}}\left( x\right) $ are the massless and massive Pauli-Jordan causal propagators, respectively. Moreover, their expressions are well known in the literature \cite{BogoIntro,Naka},%
\begin{subequations}
\begin{align}
D_{0} &=\frac{1}{2\pi }sgn\left( x_{0}\right) \delta \left(x^{2}\right),  \\
 D_{m_{a}} &=\frac{sgn\left(
x_{0}\right)}{2\pi } \left[ \delta \left( x^{2}\right) -
\frac{m_{a}\theta \left( x^{2}\right)}{2\sqrt{x^{2}}}J_{1}\left( m_{a}\sqrt{x^{2}}\right) \right] , \\
D_{0}^{\prime } &=-\frac{1}{8\pi }sgn\left( x_{0}\right) \theta \left( x^{2}\right) ,
\end{align}
\end{subequations}
where $J_{1}$ is the first-order Bessel function. Since the support
of the distribution $\delta \left( x^{2}\right) $ is contained in the surface of the backward and forward light cone,
and the support of $\theta \left( x^{2}\right) $ is contained in the closed forward light cone, this assertion is
valid for their derivatives as well. Thus, we have shown that the causal electromagnetic propagator
$D_{\mu \nu}\left( x\right) $, Eq.\eqref{1.15}, in the \emph{nonmixing} gauge has causal support, i.e.,
$\text{Supp}~\left( D_{\mu \nu }\right) \in \bar{V}^{+}\cup \bar{V}^{-}$.

\section{The Epstein-Glaser causal method}

\label{sec:2}

In order to discuss scattering processes in the framework of field theory we shall make use
of the causal framework proposed by Epstein and Glaser \cite{EpsGlas,Scharf}, a method
that explicitly uses the causal structure as a powerful tool. One of the remarkable features of this
proposal is the introduction of a test function $g$, belonging to the Schwartz space, defined
in the spacetime such that $g\left( x\right) \in \left[ 0,1\right] $. The test function plays the role of
switching the interaction in some region of the spacetime. Then, the \emph{S-matrix} is necessarily viewed as an operator-valued functional of $g$: $ S=S\left[ g\right] $.
We shall now briefly review the main points of the Epstein-Glaser causal method. \footnote{ A detailed
discussion can be found in Refs.\cite{ref43,ref49}.}

We recall that in the Epstein-Glaser approach the \emph{S-matrix} can be written in the following formal perturbative series%
\begin{align}
S\left[ g\right] =  1+\sum\limits_{n=1}^{\infty }\frac{1}{n!}\int
dx_{1}dx_{2}...dx_{n} T_{n}\left( x_{1},x_{2},...,x_{n}\right) g\left(
x_{1}\right) g\left( x_{2}\right) ...g\left( x_{n}\right) ,  \label{2.1}
\end{align}%
where we can identify the quantity $T_{n}$ as an operator-valued distribution, which is determined
inductively term by term, and $g^{\otimes n}$ is its respective test function. Moreover, we have that the test function $g$ belongs to the Schwartz space $\mathcal{J}\left( R^{4}\right) $. It should be
emphasized, however, that this formalism considers only free asymptotic fields acting on the Fock space to construct the \emph{S-matrix} $S\left[ g\right] $. For instance, for GQED (as well as for QED) we have the free electromagnetic and fermionic fields: $A_{\mu }$, $\psi$, and $\bar{\psi}$. Then for GQED, $T_{1}$ takes the following form \cite{RBufBMP}:%
\begin{equation}
T_{1}\left( x\right) =ie\colon\bar{\psi}\left( x\right) \gamma ^{\mu }\psi \left(x\right) \colon A_{\mu }\left( x\right) ,  \label{2.2}
\end{equation}%
where $"\colon$ $\colon"$ indicates the normal ordering, and (in this approach) $e$ is the normalized coupling constant. Moreover, in the Epstein-Glaser method only mathematically well-defined distributional products are introduced, such as the following intermediate $n$-point distributions%
\begin{align}
A_{n}^{\prime }\left( x_{1},...,x_{n}\right) &\equiv \sum_{P_{2}}\tilde{T}%
_{n_{1}}\left( X\right) T_{n-n_{1}}\left( Y,x_{n}\right) ,  \label{2.3} \\
R_{n}^{\prime }\left( x_{1},...,x_{n}\right) &\equiv \sum_{P_{2}}T_{n-n_{1}}\left( Y,x_{n}\right)
 \tilde{T}_{n_{1}}\left(X\right) ,  \label{2.4}
\end{align}%
where $P_{2}$ are all partitions of $\left\{ x_{1},...,x_{n-1}\right\} $ into the disjoint sets $X$, $Y$ such that $\left\vert X\right\vert=n_{1}\geq 1$ and $\left\vert Y\right\vert \leq n-2$. Moreover, if the sums in Eqs.\eqref{2.3} and \eqref{2.4} are extended over all partitions $P_{2}^{0}$, including the empty set, important distributions may be obtained. These are namely the advanced and retarded distributions%
\begin{align}
A_{n}\left( x_{1},...,x_{n}\right) &\equiv \sum_{P_{2}^{0}}\tilde{T}%
_{n_{1}}\left( X\right) T_{n-n_{1}}\left( Y,x_{n}\right),\notag \\
 &=A_{n}^{\prime }\left( x_{1},...,x_{n}\right)
+T_{n}\left(x_{1},...,x_{n}\right) , \label{2.3a}\\
R_{n}\left( x_{1},...,x_{n}\right) &\equiv \sum_{P_{2}^{0}}T_{n-n_{1}}\left( Y,x_{n}\right) \tilde{T}_{n_{1}}\left(X\right),\notag \\
&=R_{n}^{\prime }\left( x_{1},...,x_{n}\right) +T_{n}\left(
x_{1},...,x_{n}\right) . \label{2.4a}
\end{align}%
By making use of causal properties, one may conclude that $R_{n}$ and $A_{n}$ have retarded and advanced support, respectively,
\begin{subequations}
\begin{align}
\text{Supp}~R_{n}\left( x_{1},...,x_{n}\right) \subseteq \Gamma _{n-1}^{+}\left(x_{n}\right), \\
 \text{Supp}~A_{n}\left( x_{1},...,x_{n}\right) \subseteq
\Gamma _{n-1}^{-}\left( x_{n}\right) ,
\end{align}%
\end{subequations}
where $\Gamma _{n-1}^{\pm }\left( x_{n}\right) =\left\{ \left(x_{1},...,x_{n}\right) ~| ~ x_{j}\in \bar{V}^{\pm }\left(
x_{n}\right) ,~ \forall ~ j=1,\ldots,n-1\right\} $, and $\bar{V}^{\pm
}\left( x_{n}\right) $ is the closed forward (backward) cone. These two
distributions cannot be determined by the induction assumption only; in fact, they are obtained
by the \emph{splitting} process \cite{ref47} of the so-called \emph{causal} distribution, defined as
\begin{align}
D_{n}\left( x_{1},...,x_{n}\right)& \equiv R_{n}^{\prime }\left(
x_{1},...,x_{n}\right) -A_{n}^{\prime }\left( x_{1},...,x_{n}\right), \notag\\
&=R_{n}\left( x_{1},...,x_{n}\right) -A_{n}\left( x_{1},...,x_{n}\right) .
\end{align}%
In the case of the GQED we can write $D_{n}$ as it follows%
\begin{align}
D_{n}\left( x_{1},...,x_{n}\right) =\sum\limits_{k}d_{n}^{k}\left(
x_{1},...,x_{n}\right) \colon \prod\limits_{j}\bar{\psi}\left( x_{j}\right)
\prod\limits_{l}\psi \left( x_{l}\right) \prod\limits_{m}A\left(x_{m}\right) \colon,
\end{align}%
where $d_{n}^{k}\left( x_{1},...,x_{n}\right) $ is the numerical part of the \emph{causal} distribution $D_{n}$.
Moreover, by the translational invariance of $d_{n}^{k}$ one may show that it depends only on the relative coordinates:
\begin{equation}
d\left( x\right) \equiv d_{n}^{k}\left( x_{1}-x_{n},...,x_{n-1}-x_{n}\right)
\in \mathcal{J}^{\prime }\left(\mathbb{R}^{m}\right) ,\text{ \ }m=4\left( n-1\right) .
\end{equation}%
As was emphasized above, an important step in this inductive construction is the splitting process of
the \emph{causal} distribution, but its splitting at the origin $\left\{ x_{n}\right\} =\Gamma _{n-1}^{+}
\left( x_{n}\right) \cap \Gamma _{n-1}^{-}\left( x_{n}\right) $ can be translated equivalently to having its
numerical part $d$ split into the advanced and retarded distributions $a$ and $r$, respectively.
Another important point to be analyzed is the convergence of the sequence $\left\{ \left\langle d,\phi
_{\alpha }\right\rangle \right\} $, where $\phi _{\alpha }$ has decreasing support when $\alpha
\rightarrow 0^{+}$ and also belongs to the Schwartz space $\mathcal{J}$.

From the aforementioned analysis we can find some natural distributional definitions. For instance, we may define $d$
as being a distribution of \emph{singular order} $\omega $ if its Fourier transform $\hat{d}\left( p\right) $
has a quasiasymptotic $\hat{d}_{0}\left( p\right) \neq 0$ at $p=\infty $ with regard to a positive continuous
function $\rho\left( \alpha \right) $, $\alpha >0$, if the limit%
\begin{equation}
\lim_{\alpha \rightarrow 0^{+}}\rho \left( \alpha \right) \left\langle \hat{d%
}\left( \frac{p}{\alpha }\right) ,\phi \left( p\right) \right\rangle
=\left\langle \hat{d}_{0}\left( p\right) ,\phi \left( p\right) \right\rangle \neq 0 \label{2.10}
\end{equation}%
exists in $\mathcal{J}^{\prime }\left( \mathbb{R}^{m}\right) $. By the scaling transformation one may derive that the
\textit{power-counting} function $\rho \left( \alpha \right) $ satisfies
\begin{equation}
\lim_{\alpha \rightarrow 0}\frac{\rho \left( a\alpha \right) }{\rho \left(
\alpha \right) }=a^{\omega },\quad \forall ~ a>0,
\end{equation}%
with%
\begin{equation}
\rho \left( \alpha \right) \rightarrow \alpha ^{\omega }L\left( \alpha
\right) ,\text{ when }\alpha \rightarrow 0^{+},
\end{equation}%
where $L\left( \alpha \right) $ is a quasiconstant function at $\alpha =0$. Of course, there
is an equivalent definition of the above process in coordinate space, but -- since the splitting
process is more easily accomplished in momentum space -- this one suffices for our purposes.
Moreover, we specify the splitting problem by requiring that the splitting procedure must preserve
the singular order of the distributions. From these very definitions we have two distinct cases
depending on the value of $\omega $ \cite{ref49}:

(i) \emph{Regular} distributions; for $\omega <0$. In this case
the solution of the splitting problem is unique and the retarded distribution is defined
by multiplying $d$ by step functions. Its form in momentum space is
\begin{equation}
\hat{r}\left( p\right) =\frac{i}{2\pi }sgn\left( p^{0}\right) \int_{-\infty}^{+\infty }dt
\frac{\hat{d}\left( tp\right) }{\left( 1-t+sgn\left(p^{0}\right) i0^{+}\right) },  \label{2.5}
\end{equation}%
which can be identified as a dispersion relation without subtractions.

(ii) \emph{Singular} distributions; for $\omega \geq 0$. In this case the solution cannot be
obtained directly, as it was in the \emph{regular} case. But, after a careful mathematical treatment,
it may be shown that the retarded distribution is given by the so-called central splitting solution%
\begin{equation}
\hat{r}\left( p\right) =\frac{i}{2\pi }sgn\left( p^{0}\right) \int_{-\infty}^{+\infty }dt
\frac{\hat{d}\left( tp\right) }{t^{\omega +1}\left(1-t+sgn\left( p^{0}\right) i0^{+}\right) },
\end{equation}%
which can be identified as a dispersion relation with $\omega +1$ subtractions.

\section{Bhabha scattering}

\label{sec:3}

Now that we have obtained all the necessary tools and developed important ideas we can concentrate our attention on inductively determining the terms of the \emph{S-matrix} \eqref{2.1}. In particular, we are interested here in the term corresponding to Bhabha scattering. Hence, in order to accomplish this, the perturbative program is used when constructing the intermediate distributions,
\begin{equation}
R_{2}^{\prime }\left( x_{1},x_{2}\right) =-T_{1}\left( x_{2}\right)
T_{1}\left( x_{1}\right), ~~ A_{2}^{\prime }\left( x_{1},x_{2}\right)
=-T_{1}\left( x_{1}\right) T_{1}\left( x_{2}\right) ,
\end{equation}%
and subsequently when constructing the causal distribution $D_{2}$,
\begin{equation}
D_{2}\left( x_{1},x_{2}\right) =R_{2}^{\prime }\left( x_{1},x_{2}\right)
-A_{2}^{\prime }\left( x_{1},x_{2}\right) =\left[ T_{1}\left( x_{1}\right)
,T_{1}\left( x_{2}\right) \right] .  \label{3.1}
\end{equation}
For GQED we have Eq.\eqref{2.2} as the first perturbative term: $T_{1}\left(x\right) =ie\colon \bar{\psi}
\left( x\right) \gamma ^{\mu }\psi \left(x\right) \colon A_{\mu }\left( x\right) $. Hence, after applying
the Wick theorem for the normally ordered product, we obtain terms associated with the Bhabha scattering contributions:
\begin{subequations}
\begin{align}
R_{2}^{\prime}\left( x_{1},x_{2}\right) =&e^{2}\colon \bar{\psi}\left( x_{2}\right) \gamma ^{\mu } \overbrace{A_{\mu }\left( x_{2}\right) \psi \left( x_{2}\right) \bar{\psi}\left( x_{1}\right) A_{\nu }\left(
x_{1}\right) }\gamma ^{\nu }\psi \left( x_{1}\right) \colon , \\
A_{2}^{\prime}\left( x_{1},x_{2}\right) =&e^{2}\colon \bar{\psi}\left( x_{1}\right) \gamma ^{\mu }\overbrace{A_{\mu }\left( x_{1}\right) \psi \left( x_{1}\right) \bar{\psi}\left( x_{2}\right) A_{\nu }\left(
x_{2}\right) }\gamma ^{\nu }\psi \left( x_{2}\right) \colon .
\end{align}
\end{subequations}
We have that the electromagnetic contraction is
\begin{equation*}
\overbrace{A_{\mu }\left( x\right) A_{\nu }\left( y\right) }\equiv \left[
A_{\mu }^{\left( -\right) }\left( x\right) ,A_{\nu }^{\left( +\right)
}\left( y\right) \right] =iD_{\mu \nu }^{\left( +\right) }\left( x-y\right)
\end{equation*}%
where $D_{\mu \nu }^{\left( \pm \right) }\left(x\right)$ are the PF and NF parts of the electromagnetic propagator \eqref{1.14}. After some calculation, we arrive at the following expression for the \emph{causal} distribution \eqref{3.1}:
\begin{align}
D_{2}\left( x_{1},x_{2}\right) =-ie^{2}\colon \bar{\psi}\left(
x_{1}\right) \gamma ^{\mu }\psi \left( x_{1}\right)  D_{\mu \nu }\left(x_{1}-x_{2}\right) \bar{\psi}\left( x_{2}\right) \gamma ^{\nu }\psi \left(x_{2}\right) \colon ,
\end{align}%
where $D_{\mu \nu }$ is the \emph{causal} electromagnetic propagator\eqref{1.15}. Moreover,
we have shown above that this distribution has causal support: $\text{Supp}~D_{\mu \nu }\left(
x_{1},x_{2}\right) \subseteq \Gamma _{2}^{+}\left( x_{2}\right) \cup \Gamma
_{2}^{-}\left( x_{2}\right) $. Nevertheless, in order to determine the singular order of this propagator,
we shall follow the criterion in momentum space, Eq.\eqref{2.10}. Hence, from Eq.\eqref{1.16}
we write $\hat{D}_{\mu \nu }\left( \frac{k}{\alpha }\right) $ in the significant leading contribution as $\alpha \rightarrow 0^{+}$,
\begin{equation}
\hat{D}_{\mu \nu }\left( \frac{k}{\alpha }\right) \simeq \alpha ^{4}\left\{
\frac{i}{2\pi }sgn\left( k_{0}\right) \left[ m_{a}^{2}\delta
^{\left( 1\right) }\left( k^{2}\right) \right] \right\} g_{\mu \nu }.
\end{equation}%
It should be emphasized that in order to obtain the correct singular order we must consider the whole distribution, and that in obtaining the above expansion we have made use of $sgn\left( \frac{k_{0}}{\alpha }\right) =sgn\left( k_{0}\right) $, the scale property \eqref{B.4}, and the Taylor expansion \eqref{B.5} of the Dirac-$\delta $ distribution. This means that the causal propagator $\hat{D}_{\mu \nu }$ is a \emph{regular} distribution with singular order%
\begin{equation}
\omega ^{Pod}\left( \hat{D}_{\mu \nu }\right) =-4<0.
\end{equation}%
This result is more regular than the one obtained using QED \cite{Scharf}, where $%
\omega ^{Max}=-2$. Therefore, as we have determined all the necessary conditions, we are now
in position to evaluate the retarded distribution \eqref{2.4a}. For this purpose we use the
regular splitting formula \eqref{2.5} to write
\begin{align}
\hat{R}_{\mu \nu }\left( k\right)  =&\left[g_{\mu \nu }-\left( 1-\xi \right) \frac{k_{\mu }k_{\nu }}
{m_{a}^{2}}\right]\frac{i}{2\pi }sgn\left(k_{0}\right) \int\limits_{-\infty }^{+\infty }dt\frac{\left( \hat{D}%
_{0}\left( tk\right) -\hat{D}_{m_{a}}\left( tk\right) \right) }{1-t+sgn\left( k_{0}\right) i0^{+}} \notag \\
&-\left( 1-\xi \right) k_{\mu }k_{\nu }\frac{i}{2\pi }sgn\left(k_{0}\right) \int\limits_{-\infty }^{+\infty }
dt\frac{t^{2}\hat{D}_{0}^{\prime }\left( tk\right) }{1-t+sgn\left( k_{0}\right) i0^{+}}. 
\end{align}%
In order to evaluate the dispersion integrals one can make use of the explicit expression for the propagators:
$\hat{D}_{0}\left(k\right) $, $\hat{D}_{m_{a}}\left( k\right) $, and $\hat{D}_{0}^{\prime}\left( k\right) $,
Eq.\eqref{1.16a}. Finally, we find the expression for the electromagnetic \emph{retarded} propagator%
\begin{align}
\hat{R}_{\mu \nu }\left( k\right) =\left[g_{\mu \nu }-\left( 1-\xi \right) \frac{k_{\mu }k_{\nu }}{m_{a}^{2}}
\right]\left[ \hat{R}_{0}\left(k\right) -\hat{R}_{m_{a}}\left( k\right) \right]-\left( 1-\xi \right)
k_{\mu }k_{\nu }\hat{R}_{0}^{\prime }\left( k\right) ,  \label{3.2}
\end{align}%
where we have defined the following quantities for $k^{2}>0$:
\begin{subequations}
\begin{align}
\hat{R}_{0}\left( k\right)  &=-\left( 2\pi \right) ^{-2}\frac{1}{k^{2}+sgn\left( k_{0}\right) i0^{+}}, \label{3.4a}\\
\hat{R}_{m_{a}}\left( k\right)&=-\left( 2\pi \right) ^{-2}\frac{1}{k^{2}-m_{a}^{2}+sgn\left( k_{0}\right)i0^{+}},\label{3.4b}\\
\hat{R}_{0}^{\prime }\left( k\right)  &=-\left( 2\pi \right) ^{-2}\left(
\frac{1}{k^{2}+sgn\left( k_{0}\right) i0^{+}}\right) ^{2}.  \label{3.4c}
\end{align}
\end{subequations}
Although Eq.$\left( \ref{3.1}\right) $ provides several terms, we can consider only those
associated with the Bhabha scattering contribution, $T_{2}\left( x_{1},x_{2}\right) $.
This contribution is obtained from the relation:%
\begin{equation}
T_{2}\left( x_{1},x_{2}\right) =R_{2}\left( x_{1},x_{2}\right)
-R_{2}^{\prime}\left( x_{1},x_{2}\right) ,
\end{equation}%
where $R_{2}$ is the retarded part of $D_{2}$. From the previous results we have that
\begin{align}
R_{2}\left( x_{1},x_{2}\right) =-ie^{2}\colon \bar{\psi}\left(
x_{1}\right) \gamma ^{\mu }\psi \left( x_{1}\right) R_{\mu \nu }\left(
x_{1}-x_{2}\right) \bar{\psi}\left( x_{2}\right) \gamma ^{\nu }\psi \left(x_{2}\right) \colon .
\end{align}%
It then follows that the complete contribution $T_{2}$ can be written
\begin{align}
T_{2}\left( x_{1},x_{2}\right) =-ie^{2}\colon \bar{\psi}\left( x_{1}\right) \gamma ^{\mu }\psi \left( x_{1}\right)  D_{\mu \nu }^{F}\left( x_{1}-x_{2}\right) \bar{\psi}\left( x_{2}\right) \gamma ^{\nu
}\psi \left( x_{2}\right) \colon , \label{3.4}
\end{align}%
where $D_{\mu \nu }^{F}$ is defined, in the momentum space, by the relation:%
\begin{equation}
\hat{D}_{\mu \nu }^{F}\left( k\right) =\hat{R}_{\mu \nu }\left( k\right) -%
\hat{R}_{\mu \nu }^{\prime }\left( k\right).
\end{equation}%
Therefore, by replacing the expressions for $\hat{R}_{\mu \nu }$ and $\hat{R}_{\mu \nu
}^{\prime }=\hat{D}_{\mu \nu }^{\left( -\right) }$, Eqs.\eqref{3.2} and \eqref{1.14}, respectively,
we obtain the following expression for the propagator $\hat{D}_{\mu \nu }^{F}$:
\begin{align}
\hat{D}_{\mu \nu }^{F}\left( k\right) =\left[g_{\mu \nu }-\left( 1-\xi \right) \frac{k_{\mu }k_{\nu }} {m_{a}^{2}}\right]\left(\hat{D}_{0}^{F} \left( k\right)-\hat{D}_{m_{a}}^{F}\left( k\right) \right)  -\left(1-\xi \right) k_{\mu }k_{\nu }\hat{D}_{0}^{\prime F}\left( k\right), \label{3.6}
\end{align}%
with the quantities $\hat{D}_{0}^{F}$, $\hat{D}_{m_{a}}^{F}$ and $\hat{D}%
_{0}^{\prime F}$ given by \footnote{These are obtained from Eqs.\eqref{1.8}, \eqref{1.9}, \eqref{1.10} and \eqref{3.4a}-\eqref{3.4c}.}%
\begin{gather}
\hat{D}_{0}^{F} =-\left( 2\pi \right) ^{-2}\frac{1}{%
k^{2}+i0^{+}}, \quad  \hat{D}_{0}^{\prime F} =-\left( 2\pi \right) ^{-2}\left(\frac{1}{k^{2}+i0^{+}}\right) ^{2}, \quad
\hat{D}_{m_{a}}^{F}=-\left( 2\pi \right) ^{-2}%
\frac{1}{k^{2}-m_{a}^{2}+i0^{+}}. \label{3.6a}
\end{gather}%
This is the \emph{generalized} photon propagator in the \emph{nonmixing} gauge condition. Moreover, it should be emphasized that all the poles are well-defined in Eq.\eqref{3.6}, where we have used neither the Feynman
$i\epsilon $-prescription nor the Wick rotation \cite{JDBjor,BogoIntro}.

\subsection{Bhabha's Cross Section}

By definition, a transition probability $\mathcal{P}_{fi}$ is given as \cite{JDBjor,BogoIntro,Scharf}
\begin{equation}
\mathcal{P}_{fi}\equiv \left( \left\vert \psi _{f}\right\rangle ,\left\vert \psi _{i}\right\rangle \right) ^{2}
=\left\vert S_{fi}\right\vert^{2}=\left\vert \left\langle \psi _{f}\left\vert S_{2}^{\left( 1\right)
}\right\vert \psi _{i}\right\rangle \right\vert ^{2}+\cdots .
\end{equation}%
But this quantity has no meaning if it is not written as a function of
wave packets: it must also consider the states describing Bhabha scattering, with the initial state $d_{s_{i}}^{\dag }\left( \mathbf{p}_{i}\right) b_{\sigma _{i}}^{\dag }\left( \mathbf{q}_{i}\right) \left\vert \Omega
\right\rangle$ and the final state $d_{s_{f}}^{\dag }\left( \mathbf{p}_{f}\right) b_{\sigma _{f}}^{\dag }\left(\mathbf{q}_{f}\right) \left\vert \Omega \right\rangle $ written, respectively, as
\begin{subequations}
\begin{align}
\left\vert \psi _{i}\right\rangle &=\int d^{3}p_{i}d^{3}q_{1}\psi _{i}\left(
\mathbf{p}_{1},\mathbf{q}_{1}\right) d_{s_{i}}^{\dag }\left( \mathbf{p}_{1}\right) b_{\sigma _{i}}^{\dag }
\left( \mathbf{q}_{1}\right) \left\vert \Omega \right\rangle , \\
\left\vert \psi _{f}\right\rangle &=\int d^{3}p_{2}d^{3}q_{2}\psi _{f}\left(
\mathbf{p}_{2},\mathbf{q}_{2}\right) d_{s_{f}}^{\dag }\left( \mathbf{p}%
_{2}\right) b_{\sigma _{f}}^{\dag }\left( \mathbf{q}_{2}\right) \left\vert\Omega \right\rangle ,
\end{align}%
\end{subequations}
where $\left( \mathbf{q}_{i},\sigma _{i}\right)$ and $\left( \mathbf{q}_{f},\sigma _{f}\right) $ are the momentum and spin of the ingoing and outgoing electron, respectively, whereas $\left( \mathbf{p}_{i},s_{i}\right) $ and $\left( \mathbf{p}_{f},s_{f}\right) $ are the momentum and spin of the ingoing and outgoing positron, respectively. Moreover, to obtain the simplest case we shall consider a set of orthogonal wave packets, and we take into account the fact that the initial wave packets are concentrated
in $\mathbf{p}_{i}$, $\mathbf{q}_{i}$ and have fixed spins. Furthermore, if we consider the ingoing electron $\left( 1\right) $ as a target and define $\mathbf{v}$ as the relative velocity of the ingoing
positron, then, for an average cylinder of radius $R$ parallel to $\mathbf{v}$, we arrive at \cite{Mandels,Scharf}
\begin{align}
\sum\limits_{f}\mathcal{P}_{fi}\left( R\right) =\frac{1}{\pi R^{2}}\bigg[\frac{\left( 2\pi \right) ^{2}}
{\left\vert \mathbf{v}\right\vert }\int d^{3}p_{2}d^{3}q_{2}\delta \left(p_{2}+q_{2}-p_{i}-q_{i}\right)\left\vert \mathcal{M}_{s_{i}\sigma _{i}s_{f}\sigma
_{f}}\left( p_{i},q_{i},p_{2},q_{2}\right) \right\vert ^{2} \bigg] ,
\end{align}%
where $\mathcal{M}$ is a distributional quantity related to the \emph{S-matrix} by Eq.\eqref{B13}. The scattering cross section in the laboratory frame is given by \cite{Mandels,Scharf}
\begin{equation}
\sigma \equiv \lim_{R\rightarrow \infty }\pi R^{2}\sum\limits_{f}\mathcal{P}_{fi}\left( R\right) ,
\end{equation}%
which can be written in a Lorentz-invariant form as a function of the normalized electron mass $m$ and the energies of the ingoing fermions \cite{Mandels,Scharf},
\begin{align}
\sigma =\left( 2\pi \right) ^{2}\frac{E\left( p_{i}\right) E\left(q_{i}\right) }{\sqrt{\left( p_{i}q_{i}\right)
^{2}-m^{4}}}\sum\limits_{s_{f}\text{ }\sigma _{f}}\int d^{3}p_{2}d^{3}q_{2}\delta \left( p_{2}+q_{2}-p_{i}-q_{i}\right)  \left\vert \mathcal{M}%
_{s_{i}\sigma _{i}s_{f}\sigma _{f}}\left( p_{i},q_{i},p_{2},q_{2}\right)\right\vert ^{2}.
\end{align}%
For simplicity, we shall not consider the polarizations of the ingoing and outgoing fermions; hence, we
shall consider the sum over $s_{f}$ and $\sigma _{f}$, and the average over $s_{i}$ and $\sigma _{i}$.
Finally, we can write the differential cross section in the center-of-mass frame \cite{Mandels,Scharf}, \footnote{%
A detailed calculation of this quantity can be found in Appendix \ref{sec:A}.}
\begin{equation}
\frac{d\sigma }{d\Omega }=\dfrac{e^{4}}{32\left( 2\pi
\right) ^{2}E^{2}}\mathcal{F}\left( s,t,u\right) ,  \label{3.7}
\end{equation}%
with $E=E\left( p_{i}\right)$, and $s$, $t$, $u$ are the Mandelstam variables \cite{Mandels}, defined as follows%
\begin{subequations}
\begin{align}
s& =\left( p_{i}+q_{i}\right) ^{2}=\left( p_{f}+q_{f}\right)^{2} =2m^{2}+2\left( p_{f}.q_{f}\right) ,\label{3.7a}\\
t& =\left( p_{i}-p_{f}\right) ^{2}=\left( q_{i}-q_{f}\right)
^{2} =2m^{2}-2\left( q_{i}.q_{f}\right) ,\label{3.7b}\\
u& =\left( p_{f}-q_{i}\right) ^{2}=\left( p_{i}-q_{f}\right)^{2} =2m^{2}-2\left( p_{i}.q_{f}\right) . \label{3.7c}
\end{align}%
\end{subequations}
The function $\mathcal{F}\left( s,t,u\right) $ is given by Eq.\eqref{A6},%
\begin{align}
\mathcal{F}\left( s,t,u\right) =\frac{\left[ s^{2}+u^{2}+8m^{2}t-8m^{4}\right] }{t^{2}
\left( 1-\frac{t}{m_{a}^{2}}\right) ^{2}}+\frac{\left[u^{2}+t^{2}+8m^{2}s-8m^{4}\right] }{s^{2}
\left( 1-\frac{s}{m_{a}^{2}}\right) ^{2}}+2\frac{\left[ u^{2}-8m^{2}u+12m^{4}\right] }
{st\left( 1-\frac{s}{m_{a}^{2}}\right) \left( 1-\frac{t}{m_{a}^{2}}\right) }. \label{3.9a}
\end{align}%
Furthermore, we see that when $m_{a}\rightarrow \infty $ this expression reduces to the QED one \cite{Scharf},
a fact that is in accordance with the relation between the Podolsky to Maxwell theories. Therefore, the 
Podolsky mass is larger than the electron mass $m=0,510~MeV$; for instance, in Ref.~\cite{RBufBMP} it was obtained that $m_{a}\geq 37,59~GeV$. Hence, we can conclude that in the nonrelativistic and lower-energy regime the differential cross sections are the same ones as those in QED  \cite{Mandels}. On the other hand, this indicates that GQED effects must be considered in the so-called high-energy regime,
\begin{equation}
m^{2}\ll s \sim \left\vert t\right\vert \sim \left\vert u\right\vert.
\end{equation}
Thus, we conclude that the terms associated to the electron mass in Eq.\eqref{3.9a} can be neglected. Thus, the differential cross section \eqref{3.7} in the high-energy regime is given by \footnote{In this expression we introduced the fine-structure constant, $\alpha=e^2/4\pi $, and $\sqrt{s}=2E$ is the center-of-mass energy; see Eq.\eqref{3.10}.}
\begin{align}
\frac{d\sigma }{d\Omega }=\frac{\alpha^{2}}{2s}\bigg[\frac{s^{2}+u^{2} }{t^{2}\left( 1-\frac{t}{m_{a}^{2}}\right) ^{2}}+\frac{u^{2}+t^{2} }{s^{2}\left( 1-\frac{s}{m_{a}^{2}}\right) ^{2}}+2\frac{ u^{2} }{st\left( 1-\frac{s}{m_{a}^{2}}\right) \left( 1-\frac{t}{m_{a}^{2}}\right) }\bigg].  \label{3.9d}
\end{align}
This relation is identical to the phenomenological formula of Bhabha scattering. Hence, we can identify the
free parameter $m_a$ as being related to the phenomenological cutoff parameter $\Lambda_+$ \cite{ref19}. In Ref.~\cite{ref21} (TASSO collaboration) measurements of the differential cross sections were presented at $95\%$ C.L. and
at total central-of-mass energy $12~GeV\leq\sqrt{s}\leq46,8~GeV$, which results in the value $m_a\geq 370~GeV$;
other experiment measurements \cite{ref23} led to values of the same order of magnitude for $m_a$.

For practical reasons we also consider the high-energy regime below the Podolsky mass
\begin{equation}
m^{2}\ll s \sim \left\vert t\right\vert \sim \left\vert u\right\vert <m_{a}^{2},
\end{equation}
thus, in the leading-order term in $\frac{\sqrt{s}}{m_{a}}$, Eq.\eqref{3.9a} can be written as
\begin{align}
\mathcal{F}\left( s,t,u\right)\approx  \left( \frac{s^{2}+u^{2}}{t^{2}}+%
\frac{u^{2}+t^{2}}{s^{2}}+2\frac{u^{2}}{st}\right) +2\frac{1}{m_{a}^{2}}%
\left( \frac{s^{2}+2u^{2}}{t}+\frac{2u^{2}+t^{2}}{s}\right) .  \label{b3}
\end{align}%
Moreover, the differential cross section is conventionally evaluated in the center-of-mass frame, where we have the relations
\begin{equation}
\mathbf{p}_{i}=-\mathbf{q}_{i}\equiv \mathbf{p}, ~~  \mathbf{p}_{f}=-\mathbf{q}_{f},
~~ E\left(p_{i}\right) =E\left( q_{i}\right) \equiv E, ~~ E\left( p_{f}\right)=E\left( q_{f}\right).
\end{equation}
Besides, from energy-momentum conservation it follows that:
$ E\left( p_{i}\right) =E\left( p_{f}\right) \rightarrow \left\vert \mathbf{p}_{f}\right\vert =\left\vert \mathbf{p}_{i}\right\vert $,
and by defining $\theta $ as the center-of-mass scattering angle the Mandelstam variables \eqref{3.7a}-\eqref{3.7c} read
\begin{equation}
s=4m^{2}+4\mathbf{p}^{2}=4E^{2},\quad u=-4\mathbf{p}^{2}\cos ^{2}\frac{%
\theta }{2},\quad t=-4\mathbf{p}^{2}\sin ^{2}\frac{\theta }{2}.  \label{3.10}
\end{equation}%
Finally, after some manipulations, we obtain the following outcome for the differential cross section:
\begin{align}
\frac{d\sigma }{d\Omega }=\frac{\alpha ^{2}}{256E^{2}}\frac{\left( \cos 2\theta +7\right)
^{2}}{\sin ^{4}\frac{\theta}{2} }-\frac{\alpha ^{2}}{32m_{a}^{2}}\allowbreak \frac{\left( 45\cos \theta +6\cos
2\theta +3\cos 3\theta +42\right) }{\sin ^{2}\frac{\theta}{2} }. \label{4.6}
\end{align}%
This expression adds an additional term to the usual ultrarelativistic limit of the Bhabha formula,
which we shall call the \emph{GQED correction to the ultrarelativistic limit of the Bhabha formula}.

Furthermore, we can identify the first term in Eq.\eqref{4.6} as the QED contribution and the second as the GQED
correction. Thus, the GQED correction to Bhabha scattering may be evaluated using the formula
\begin{align}
\delta  =\left( \frac{d\sigma }{d\Omega }\right)^{GQED}  \left/ \left( \frac{d\sigma }{d\Omega }\right)\right. ^{QED}-1.
\end{align}
Thus, we can obtain the expression
\begin{align}
\delta =-2\left( \frac{\sqrt{s}}{m_{a}}\right) ^{2}\frac{\left( \sin ^{2}\frac{\theta}{2} \right) \left( 45\cos \theta
+6\cos 2\theta +3\cos 3\theta+42\right) }{\left( \cos 2\theta +7\right) ^{2}} . \label{4.7}
\end{align}

On the other hand, as aforementioned, the study of Bhabha scattering is relevant mainly because it is the
process employed in determining the luminosity $L$ at $e^{-}e^{+}$ colliders; in fact, we have $L=N_{Bha}/\sigma _{th}$,
where $N_{Bha}$ is the rate of Bhabha events and $\sigma _{th}$ is the Bhabha scattering cross section obtained
by theoretical calculation \cite{lum}. There are two kinematical regions of interest for the luminosity measurements: one
is the small-angle Bhabha (SABh) process, which is found at scattering angles below $6^\circ $, and is mainly dominated by
the $t$-channel photon exchange; the other is the large-angle Bhabha (LABh) process, which is found at scattering angles
above $6^\circ$, and receives important contributions from various $s$-channel (annihilation) exchanges. The SABh process
$e^{-}e^{+}\rightarrow e^{-}e^{+}$ is employed in determining the luminosity and, hence, the absolute normalization
of the cross section expression for all other $e^{-}e^{+}$ collisions. Moreover, since the luminosity is dominated by
photon exchange its contribution is calculable, in principle, by perturbative QED with arbitrary precision. Thus,
in the GQED approach we can calculate the lowest effect on this theory to the SABh process. Thus, by expanding Eq.\eqref{4.7} for small angles $\theta \ll 1$, we have that the correction is
\begin{equation}
\delta =-\left( \frac{\sqrt{s}}{m_{a}}\right) ^{2}\frac{3}{4}\theta ^{2}.
\end{equation}
Besides, we can rewrite Eq.\eqref{4.6} for the differential cross section at small angles,
\begin{equation}
\frac{d\sigma }{d\Omega }=\dfrac{4\alpha ^{2}}{%
E^{2}\theta ^{4}}\left\{ 1-\left[ \frac{1}{2}+3\left( \frac{E}{m_{a}}\right)
^{2}\right] \theta ^{2}+\cdots \right\} .
\end{equation}%
Thus, we can see that the GQED deviation for the luminosity decreases at the second-order
term the usual QED contribution.

As an additional remark, note that at the International Linear Collider (ILC) \cite{ref70}, one can reach part-per-mil accuracy with regards to the theoretical calculation, $|\delta|\leq0.1\%$. Also, the ILC operates at different center-of-mass energies, running, in principle, up to $\sqrt{s}=500~GeV$. In particular, we can calculate the GQED correction at the energy corresponding to the $ Z $ resonance, at $\sqrt{s}=91~GeV$, for the SABh process when considering an angle as small as $5^\circ $; hence, using the previous result for the lower Podolsky mass $m_a = 370~GeV$, we can estimate the GQED correction to be on order of $\delta =-0.035\%$, a value that is within the expectations of high-precision luminosity measurements.

Moreover, for the LABh process, we see from Eq.\eqref{4.7} that the GQED correction increases when the angle increases, taking its maximum value at $\theta =90^{0}$. Hence, in this case Eq.\eqref{4.7} takes the form
\begin{equation}
\delta =-\left( \frac{\sqrt{s}}{m_{a}}\right) ^{2}.\label{4.8}
\end{equation}
Although the QED contributions dominate the radiative corrections to the LABh scattering at intermediate center-of-mass energies $( 1-10GeV )$, we may still have the presence of contributions from GQED corrections in this region. For instance, from Eq.\eqref{4.8} at $\sqrt{s}=10~GeV$ and with a Podolsky mass $m_a = 370~GeV$, we can roughly estimate the GQED correction to be on the order of $\delta =-0.073\%$, a value that is within the expectations of high-precision luminosity measurements \cite{ref71}. For higher center-of-mass energies the electroweak corrections, such as the Z-boson exchange, begin to be relevant, so a more careful analysis must be made.

\section{Concluding remarks}

\label{sec:5}

In this paper we have discussed Bhabha scattering in the framework of generalized quantum electrodynamics.
The theory was quantized in the framework of the causal method of Epstein and Glaser, where this perturbative
program gave us consistent results with regards to general physical requirements, such as causality, as well as mathematically well-defined quantities, as it is embedded in the realm of distribution theory.

This approach takes into account asymptotically free conditions, and thus the the Dirac equation is used for fermionic particles and the Podolsky free field equation is used for the photon. Moreover, in the latter we
considered the so-called nonmixing gauge condition. With this gauge condition we were
able to find a suitable expression for the free electromagnetic propagator that clarified its physical content. Also, since the physical result of the transition amplitude of a given scattering process is not affected by the particular value of the gauge-fixing parameter $\xi$, we have considered $\xi =1$ in our calculations.

In our analysis of Bhabha scattering we found that our "high-energy" formula has the same form as the
phenomenological formula, which considers the cutoff parameter or Feynman regulator $\Lambda _{+}$. Thus by
identifying the Podolsky mass as this cutoff parameter, we were able to find a bound for the free
parameter $m_{a}$. Hence, from electron-positron scattering with $12~GeV \leq \sqrt{s}\leq 46,8~GeV$, it follows
that $m_{a}>370~GeV$. Moreover, from this result we can estimate the GQED corrections to the luminosity: for the
SABh process at $\theta =5^0$ and $\sqrt{s}=91~GeV$ the correction $-0.035\%$, and for the LABh process at at $\theta =90^0$ and $\sqrt{s}=10~GeV$ the correction is $-0.073\%$. These results are in accordance with the expected high-precision measurements of the future ILC. When we compare our result with the value of the Podolsky mass obtained in Ref. \cite{RBufBMP}, $m_{a}\geq 37.59 ~GeV$, we see that even this previous result was obtained by considering
experimental data of the electron's anomalous magnetic moment; this previous result is one order lower.
This support the idea that an electron-positron collider is an excellent experiment to study new particle physics.

In addition, there are many other processes that may receive contributions from GQED, in particular, since at high energies, photon and Z-boson contributions are of the same order of magnitude, we may argue that GQED could also provide a contribution to the neutral-current Drell-Yan process \cite{ref50} (being presently measured at the LHC with high precision). Evidently, the calculation of the GQED contribution to the Drell-Yan process is very similar to that used in the Bhabha case.

We believe that generalized quantum electrodynamics stands as a reasonable leptonic-photon interacting theory
and, moreover, it can cope with many "high energy" deviations from experimental results, by introducing the Podolsky mass.
Once we have developed all of the principal ideas of the causal inductive program, given in the Sec.\ref{sec:1} and \ref{sec:2}, we are in a position to perform some analyses. (i) A first step in this direction may be the explicit computation and discussion of the other second-order terms, in particular, the GQED one-loop radiative corrections, such as the vacuum polarization, the fermion self-energy, and the three-point vertex function. For this purpose, we will make use of the full strength of the Epstein-Glaser causal approach, which will lead naturally to well-defined and regularized quantities. (ii) On the other hand, we may study some other physical properties which were not considered as part of the constructed axioms, like the discrete symmetries: parity, time-reversal and charge conjugation. In particular, we want to examine the GQED (re)normalizability and gauge invariance within the perturbative causal approach \cite{ref51}, i.e., to show that the Ward-Takahashi-Fradkin identities are satisfied perturbatively order by order. These issues and others will be further elaborated, investigated and reported elsewhere.

\subsection*{Acknowledgments}

The authors would like to thanks the anonymous referee for his/her comments and suggestions to improve this paper. R.B. thanks FAPESP for full support, B.M.P. thanks CNPq and CAPES for partial support and D.E.S. thanks CNPq for full support.

\appendix

\section{Dirac's delta distribution properties}

\label{sec:B}

In this appendix we summarize some properties of the Dirac $\delta$-distribution \cite{Gwais}. If $\varphi $ is
a test function, the definition for the Dirac $\delta$-translated distribution%
\begin{equation}
\varphi \left( \pm \omega _{m}\right) =\left\langle \delta \left( k_{0}\mp
\omega _{m}\right) ,\varphi \left( k_{0}\right) \right\rangle .  \label{B.1}
\end{equation}%
Moreover, from this definition we may also obtain its derivative, $n=0,1$:%
\begin{equation}
\varphi ^{\left( n\right) }\left( \pm \omega _{0}\right) =\left( -1\right) ^{\left( n\right) }\left\langle
\delta ^{\left( n\right) }\left(k_{0}\mp \omega _{0}\right) ,\varphi \left( k_{0}\right) \right\rangle.
\label{B.2}
\end{equation}%
Another important relation involves
\begin{equation}
\theta \left( \pm \alpha \right) \delta ^{\left( 1\right) }\left( \alpha^{2}-\beta ^{2}\right) =\pm \sum
\limits_{j=0}^{1}\frac{\left( 1+j\right) !}{\left( 1-j\right) !}\left[ \frac{\delta ^{\left( 1-j\right) }
\left( \alpha\mp \beta \right) }{\left( \pm 2\beta \right) ^{2+j}}\right], \label{B.3}
\end{equation}%
with $\beta >0$. In particular, we have
\begin{equation}
\theta \left( \pm k_{0}\right) \delta \left( k_{0}^{2}-\omega
_{m}^{2}\right) =\frac{\delta \left( k_{0}\mp \omega _{m}\right) }{2\omega _{m}}.
\end{equation}
Two important identities of the Dirac $\delta $-distribution are its scale property and its Taylor expansion, respectively,
\begin{align}
\delta ^{\left( n\right) }\left( \frac{x}{\beta }\right) &=\left\vert \beta
\right\vert \beta ^{n}\delta ^{\left( n\right) }\left( x\right), \quad n=0,1,\ldots  \label{B.4} \\
\delta \left( x-\beta \right) &=\sum\limits_{n=0}^{\infty }\frac{\left(
-1\right) ^{n}}{n!}\delta ^{\left( n\right) }\left( x\right) \beta ^{n}. \label{B.5}
\end{align}

\section{Transition amplitude for Podolsky's photon exchange}

\label{sec:A}

To calculate the matrix amplitude for Bhabha scattering, we shall consider the following initial
state $\left\vert \psi _{i}\right\rangle $ and final state $\left\vert \psi _{f}\right\rangle $%
\begin{align}
\left\vert \psi _{i}\right\rangle & =d_{s_{i}}^{\dag }\left( \mathbf{p}_{i}\right) b_{\sigma _{i}}^{\dag }
\left( \mathbf{q}_{i}\right) \left\vert\Omega \right\rangle , \\
\left\vert \psi _{f}\right\rangle & =d_{s_{f}}^{\dag }\left( \mathbf{p}_{f}\right) b_{\sigma _{f}}^{\dag }
\left( \mathbf{q}_{f}\right) \left\vert\Omega \right\rangle ,
\end{align}%
where $\left( \mathbf{q}_{i},\sigma _{i}\right) $and $\left( \mathbf{q}_{f},\sigma _{f}\right) $ are the
momentum and spin of the ingoing and outgoing electron, respectively, and
$\left( \mathbf{p}_{i},s_{i}\right) $ and $\left( \mathbf{p}_{f},s_{f}\right) $ are the momentum and spin
of the ingoing and outgoing positron, respectively. Thus, the transition amplitude at the second order in the coupling
constant, in the causal approach, is given by the expression:%
\begin{equation}
S_{fi}=\left\langle \psi _{f}\left\vert S_{2}\right\vert \psi_{i}\right\rangle =\frac{1}{2!}\left\langle
\psi _{f}\left\vert \int d^{4}x_{1}d^{4}x_{2}T_{2}\left( x_{1},x_{2}\right) \right\vert \psi_{i}\right\rangle .
\end{equation}%
By substituting the expression for $T_{2}$ from Eq.\eqref{3.4} into the above result we have that%
\begin{align}
S_{fi}=-\frac{ie^{2}}{2!}\int d^{4}x_{1}d^{4}x_{2}\bigg\langle \psi_{f}\bigg\vert \colon \bar{\psi}
\left( x_{1}\right) \gamma ^{\mu }\psi\left( x_{1}\right) D_{\mu \nu }^{F}\left( x_{1}-x_{2}\right)
\bar{\psi}\left( x_{2}\right) \gamma ^{\nu }\psi \left( x_{2}\right) \colon
\bigg\vert \psi _{i}\bigg\rangle .  \label{A1}
\end{align}%
We have that the Dirac field free solutions are given by \cite{JDBjor}%
\begin{subequations}
\begin{align}
\psi \left( x\right)  =& \sum\limits_{s} \int \frac{d^{3}p}{\left( 2\pi \right) ^{\frac{3}{2}}}\bigg[b_{s}\left( \mathbf{p}\right)
u_{s}\left( \mathbf{p}\right)e^{-ipx}+d_{s}^{\dag }\left( \mathbf{p}\right) v_{s}\left(
\mathbf{p}\right) e^{ipx}\bigg] , \\
\bar{\psi}\left( x\right) =& \sum\limits_{s} \int \frac{d^{3}p}{\left( 2\pi \right) ^{\frac{3}{2}}}\bigg[b_{s}^{\dag } \left( \mathbf{p}\right) \bar{u}_{s}\left( \mathbf{p}\right) e^{ipx}+d_{s} \left(\mathbf{p}\right) \bar{v}_{s}\left( \mathbf{p}\right) e^{-ipx}\bigg],
\end{align}%
\end{subequations}
where the pair of operators $b_{s_{1}}$, $b_{s}^{\dag }$ and $d_{s_{1}}$, $d_{s}^{\dag }$ satisfy the anticommutation relations:
$\left\{b_{s_{1}}\left( \mathbf{p}_{1}\right) ,b_{s}^{\dag }\left( \mathbf{p}\right)\right\} =\delta _{s_{1}s}
\delta \left( \mathbf{p}_{1}-\mathbf{p}\right)=\left\{ d_{s_{1}}\left( \mathbf{p}_{1}\right) ,d_{s}^{\dag }\left(\mathbf{p}\right) \right\}  $,
while any other anticommutation relations vanish. Hence, we can obtain%
\begin{subequations}
\begin{align}
\psi \left( x\right) b_{s}^{\dag }\left( \mathbf{p}\right) \left\vert \Omega
\right\rangle & =\left\vert \Omega \right\rangle \left( 2\pi \right) ^{-3/2}u_{s}\left( \mathbf{p}\right) e^{-ipx}, \\
\bar{\psi}\left( x\right) d_{s}^{\dag }\left( \mathbf{p}\right) \left\vert\Omega \right\rangle &
=\left\vert \Omega \right\rangle \left( 2\pi \right) ^{-3/2}\bar{v}_{s}\left( \mathbf{p}\right) e^{-ipx}, \\
\left\langle \Omega \right\vert b_{s}\left( \mathbf{p}\right) \bar{\psi}\left( x\right) & =\left( 2\pi \right)
^{-3/2}\bar{u}_{s}\left( \mathbf{p}\right) e^{ipx}\left\langle \Omega \right\vert , \\
\left\langle \Omega \right\vert d_{s}\left( \mathbf{p}\right) \psi \left(
x\right) & =\left( 2\pi \right) ^{-3/2}v_{s}\left( \mathbf{p}\right)e^{ipx}\left\langle \Omega \right\vert ,
\end{align}%
\end{subequations}
where $\left\vert \Omega \right\rangle $ is the vacuum state and $u_{s}$ and $v_{s}$ are the positive- and
negative-energy Dirac spinors, respectively. Thus, after some algebraic manipulation, the transition amplitude \eqref{A1} can be written as:
\begin{align}
S_{fi} &=-ie^{2}\left( 2\pi \right) ^{-6}\int d^{4}x_{1}d^{4}x_{2}D_{\mu\nu }^{F}\left( x_{1}-x_{2}\right) e^{-i p_{i} x_{1}+i q_{f} x_{2}} \notag \\
&\times \bigg[ \bar{v}_{s_{i}}\left( \mathbf{p}_{i}\right) \gamma ^{\mu }v_{s_{f}}\left( \mathbf{p}_{f}\right)
\bar{u}_{\sigma _{f}}\left( \mathbf{q}_{f}\right) \gamma ^{\nu }u_{\sigma _{i}}\left( \mathbf{q}_{i}\right)
e^{i p_{f} x_{1}-i q_{i} x_{2}} \notag \\
&-\bar{v}_{s_{i}}\left( \mathbf{p}_{i}\right) \gamma ^{\mu}u_{\sigma _{i}}\left( \mathbf{q}_{i}\right)
\bar{u}_{\sigma _{f}}\left(\mathbf{q}_{f}\right) \gamma ^{\nu }v_{s_{f}}\left( \mathbf{p}_{f}\right)
e^{-i q_{i} x_{1}+i p_{f} x_{2}}\bigg].
\end{align}%
Moreover, by means of the distributional property,
\begin{equation}
\int d^{4}x_{1}d^{4}x_{2}D_{\mu \nu }^{F}\left( x_{1}-x_{2}\right)e^{ipx_{1}+iqx_{2}}=\left( 2\pi \right)
^{6}\delta \left( p+q\right) \hat{D}_{\mu \nu }^{F}\left( p\right) ,
\end{equation}%
we finally can write the transition amplitude in the short form:%
\begin{equation}
S_{fi}=\delta \left( p_{f}+q_{f}-p_{i}-q_{i}\right) \mathcal{M}, \label{B13}
\end{equation}%
where $\mathcal{M}$ is the matrix amplitude for the Podolsky photon exchange, which is explicitly defined as
\begin{align}
i\mathcal{M} &=e^{2}\bigg[ \hat{D}_{\mu \nu }^{F}\left( p_{f}-p_{i}\right)
\bar{v}_{s_{i}}\left( \mathbf{p}_{i}\right) \gamma ^{\mu }v_{s_{f}}\left(
\mathbf{p}_{f}\right) \bar{u}_{\sigma _{f}}\left( \mathbf{q}_{f}\right)
\gamma ^{\nu }u_{\sigma _{i}}\left( \mathbf{q}_{i}\right) \notag\\
&-\hat{D}_{\mu \nu }^{F}\left( -p_{i}-q_{i}\right) \bar{v}%
_{s_{i}}\left( \mathbf{p}_{i}\right) \gamma ^{\mu }u_{\sigma _{i}}\left(
\mathbf{q}_{i}\right) \bar{u}_{\sigma _{f}}\left( \mathbf{q}_{f}\right)
\gamma ^{\nu }v_{s_{f}}\left( \mathbf{p}_{f}\right) \bigg]. \label{B14}
\end{align}%
As is easily seen, the transition and matrix amplitude are both distributions, and thus they only have
meaning when wave-packets are considered.

In the calculation of the Bhabha scattering cross section, at Sect. \ref{sec:3}, we consider the
sum over the final spins and the average over the initial spins; thus, we had make the substitution $\left\vert \mathcal{M}\right\vert ^{2}\rightarrow \frac{1}{4}\sum\limits_{s_{i},\sigma _{i}}\sum\limits_{s_{f},\sigma _{f}}\left\vert \mathcal{M}\right\vert ^{2}$. Hence, from Eq.\eqref{B14} we can evaluate the quantity:%
\begin{align}
\left\vert \mathcal{M}\right\vert ^{2} =&e^{4}\bigg[ \hat{D}_{\mu \nu}^{\ast F}\left( p_{f}-p_{i}\right)
\bar{u}_{\sigma _{i}}\left( \mathbf{q}_{i}\right) \gamma ^{\nu }u_{\sigma _{f}}\left( \mathbf{q}_{f}\right)
\bar{v}_{s_{f}}\left( \mathbf{p}_{f}\right) \gamma ^{\mu }v_{s_{i}}\left( \mathbf{p}_{i}\right) \notag \\
&-\hat{D}_{\mu \nu }^{\ast F}\left( -p_{i}-q_{i}\right) \bar{v}_{s_{f}}\left( \mathbf{p}_{f}\right)
\gamma ^{\nu }u_{\sigma _{f}}\left(\mathbf{q}_{f}\right) \bar{u}_{\sigma _{i}}\left( \mathbf{q}_{i}\right)
\gamma ^{\mu }v_{s_{i}}\left( \mathbf{p}_{i}\right) \bigg]  \notag \\
&\times \bigg[ \hat{D}_{\alpha \beta }^{F}\left( p_{f}-p_{i}\right) \bar{v}_{s_{i}}\left( \mathbf{p}_{i}\right)
\gamma ^{\alpha }v_{s_{f}}\left(\mathbf{p}_{f}\right) \bar{u}_{\sigma _{f}}\left( \mathbf{q}_{f}\right) \gamma ^{\beta }u_{\sigma _{i}}\left( \mathbf{q}_{i}\right) \notag \\
&-\hat{D}_{\alpha \beta }^{F}\left( -p_{i}-q_{i}\right) \bar{v}_{s_{i}}\left( \mathbf{p}_{i}\right) \gamma ^{\alpha }u_{\sigma _{i}}\left(\mathbf{q}_{i}\right) \bar{u}_{\sigma _{f}}\left( \mathbf{q}_{f}\right) \gamma ^{\beta }v_{s_{f}}\left( \mathbf{p}_{f}\right) \bigg] .
\end{align}%
Nevertheless, in order to calculate $\frac{1}{4}\sum\limits_{s_{i},\sigma _{i}}\sum\limits_{s_{f},
\sigma _{f}}\left\vert M\right\vert ^{2}$ we can use the completeness relations for the four-spinor \cite{JDBjor}, $\sum\limits_{s}u_{s}\left( \mathbf{p}\right) \bar{u}_{s}\left( \mathbf{p}%
\right)  =\frac{\gamma .p+m}{2E}$ and $\sum\limits_{s}v_{s}\left( \mathbf{p}\right) \bar{v}_{s}\left( \mathbf{p} \right)  =\frac{\gamma .p-m}{2E}$; thus, after some calculation we obtain that%
\begin{align}
\frac{1}{4}\sum\limits_{s_{i},\sigma _{i}}\sum\limits_{s_{f},\sigma_{f}}\left\vert \mathcal{M}\right\vert ^{2}
=&\frac{e^{4}}{64E\left(p_{i}\right) E\left( q_{i}\right) E\left( p_{f}\right) E\left( q_{f}\right) } \\
&\times \bigg\{-\hat{D}_{\mu \nu }^{\ast F}\left( p_{f}-p_{i}\right) \hat{D}_{\alpha \beta }^{F}\left( -p_{i}-q_{i}\right)
\Xi _{2}^{\nu \beta \mu\alpha }\left( q_{f},p_{f},p_{i},q_{i}\right) -\left( p_{f}\rightleftarrows -q_{i}\right)    \notag \\
&+\hat{D}_{\mu \nu }^{\ast F}\left( p_{f}-p_{i}\right) \hat{D}_{\alpha \beta }^{F}
\left( p_{f}-p_{i}\right) \Xi ^{\beta \nu }\left(q_{i},q_{f},m\right) \Xi ^{\mu \alpha }\left( p_{i},p_{f},-m\right)
+\left(p_{f}\rightleftarrows -q_{i}\right) \bigg\},\notag
\end{align}%
where $E$ indicates the energy of the corresponding fermion, and we have defined the quantities
\begin{subequations}
\begin{align}
\Xi ^{\beta \nu }\left( p,q,m\right) & =tr\left[ \gamma ^{\beta }\left(
\gamma .p+m\right) \gamma ^{\nu }\left( \gamma .q+m\right) \right] , \\
\Xi _{2}^{\nu \beta \mu \alpha }\left( q_{f},p_{f},p_{i},q_{i}\right) & =tr
\left[ \gamma ^{\nu }\left( \gamma .q_{f}+m\right) \gamma ^{\beta }\left(
\gamma .p_{f}-m\right) \gamma ^{\mu }\left( \gamma .p_{i}-m\right) \gamma
^{\alpha }\left( \gamma .q_{i}+m\right) \right] .
\end{align}%
\end{subequations}
Since the longitudinal part of the photon propagator $D_{\mu \nu }^{F}$, Eq.\eqref{3.6}, does not contribute
to the transition amplitude, we can choose, without loss of generality, the gauge-fixing parameter to be $\xi =1$.
Also, by using the properties of the $\gamma $-matrices we can evaluate:%
\begin{subequations}
\begin{align}
g_{\mu \nu }g_{\alpha \beta }\Xi ^{\beta \nu }\left( q_{i},q_{f},m\right)
\Xi ^{\mu \alpha }\left( p_{i},p_{f},-m\right) & =8\left[s^{2}+u^{2}+8m^{2}t-8m^{4}\right] , \\
g_{\mu \nu }g_{\alpha \beta }\Xi _{2}^{\nu \beta \mu \alpha }\left(
q_{f},p_{f},p_{i},q_{i}\right) & =8\left[ -u^{2}+8m^{2}u-12m^{4}\right] ,
\end{align}%
\end{subequations}
where $s$, $t$, $u$ are the Mandelstam variables \cite{Mandels}. Finally, with the above results, we find that%
\begin{align}
\frac{1}{4}\sum\limits_{s_{i},\sigma _{i}}\sum\limits_{s_{f},\sigma _{f}}\left\vert \mathcal{M}\right\vert ^{2}
=&\frac{e^{4}}{8E\left(p_{i}\right) E\left( q_{i}\right) E\left( p_{f}\right) E\left( q_{f}\right) }  \label{A5} \\
&\times \bigg\{\left[ s^{2}+u^{2}+8m^{2}t-8m^{4}\right] \left[ \hat{D}_{0}^{F}-\hat{D}_{m_{a}}^{F}\right]
^{\ast }\left( \sqrt{t}\right) \left[\hat{D}_{0}^{F}-\hat{D}_{m_{a}}^{F}\right] \left( \sqrt{t}\right) +\left(
s\rightleftarrows t\right)   \notag \\
& \quad +\left[ u^{2}-8m^{2}u+12m^{4}\right] \left[ \hat{D}_{0}^{F}-\hat{D}_{m_{a}}^{F}\right] ^{\ast }
\left( \sqrt{t}\right) \left[ \hat{D}_{0}^{F}-\hat{D}_{m_{a}}^{F}\right] \left( \sqrt{s}\right)
+\left( s\rightleftarrows t\right) \bigg\}. \notag
\end{align}%
Nevertheless, from the definition of the massless and massive propagators, Eq.\eqref{3.6a}, we obtain the relations:%
\begin{subequations}
\begin{align}
\left[ \hat{D}_{0}^{F}-\hat{D}_{m_{a}}^{F}\right] ^{\ast }\left( k\right) \left[ \hat{D}_{0}^{F}-\hat{D}_{m_{a}}
^{F}\right] \left( k\right)  =&\left(2\pi \right) ^{-4}\frac{2}{k^{4}\left( 1-\frac{k^{2}}{m_{a}^{2}}\right) ^{2}}, \\
\left[ \hat{D}_{0}^{F}-\hat{D}_{m_{a}}^{F}\right] ^{\ast }\left( k\right)\left[ \hat{D}_{0}^{F}-\hat{D}_{m_{a}}
^{F}\right] \left( q\right) +\left(q\rightleftarrows k\right)  =&\left( 2\pi \right) ^{-4}\frac{1}{%
k^{2}q^{2}\left( 1-\frac{k^{2}}{m_{a}^{2}}\right) \left( 1-\frac{q^{2}}{m_{a}^{2}}\right) }.
\end{align}%
\end{subequations}
Finally, substituting the above relations in \eqref{A5} and after some simplifications, we obtain that:%
\begin{align}
\mathcal{F}\left( s,t,u\right) \equiv& \frac{8 \pi ^{4}s^{2}}{e^{4}}\left[ \frac{1}{4}\sum\limits_{s_{i},\sigma _{i}}\sum\limits_{s_{f},\sigma _{f}}\left\vert \mathcal{M}\right\vert ^{2}\right]   \notag \\
=&\frac{\left[ s^{2}+u^{2}+8m^{2}t-8m^{4}\right] }{t^{2}\left( 1-\frac{t}{m_{a}^{2}}\right) ^{2}}+
\frac{\left[ u^{2}+t^{2}+8m^{2}s-8m^{4}\right]}{s^{2}\left( 1-\frac{s}{m_{a}^{2}}\right) ^{2}}
+2\frac{\left[u^{2}-8m^{2}u+12m^{4}\right] }{st\left( 1-\frac{s}{m_{a}^{2}}\right) \left(1-\frac{t}{m_{a}^{2}}\right) }.  \label{A6}
\end{align}%


\end{document}